\documentclass[aps,preprint,showpacs,preprintnumbers,amsmath,amssymb]{revtex4}
\usepackage{amsmath,mathrsfs,amsbsy,color,graphicx,bm,amsthm,amsfonts}
\usepackage{units}
\usepackage{bbm}
\usepackage{times}
\usepackage{dcolumn}
\usepackage{mathrsfs}
\usepackage{amsmath,amssymb,epsfig}

\begin{document}

\title{Fermionic steering is not nonlocal in the background of dilaton black hole}
\author{Shu-Min Wu\footnote{smwu@lnnu.edu.cn }, Jin-Xuan Li, Xiao-Ying Jiang, Xiao-Wei Teng,  Xiao-Li Huang\footnote{ huangxiaoli1982@foxmail.com (corresponding author)}, Jianbo Lu\footnote{  lvjianbo819@163.com (corresponding author)} }
\affiliation{Department of Physics, Liaoning Normal University, Dalian 116029, China
}


\begin{abstract}
 We study the redistribution of the fermionic steering and the relation among fermionic Bell nonlocality, steering, and entanglement in the background of the Garfinkle-Horowitz-Strominger dilaton black hole.
We analyze the meaning of the fermionic steering in terms of the Bell inequality in curved spacetime.
We find that the fermionic steering, which is previously found to survive in the extreme dilaton black hole, cannot be considered to be nonlocal.
We also find that the dilaton gravity can redistribute the fermionic steering, but cannot redistribute Bell nonlocality, which means that the physically inaccessible steering is also not nonlocal.
Unlike the inaccessible entanglement,  the inaccessible steering
may increase non-monotonically with the dilaton.
Furthermore, we obtain some monogamy relations between the fermionic steering and entanglement in dilaton spacetime.
In addition, we show the difference between the fermionic and bosonic steering in curved spacetime.
\end{abstract}

\vspace*{0.5cm}
\maketitle
\section{Introduction}

Einstein-Podolsky-Rosen (EPR) steering, first discussed by Schr\"{o}dinger \cite{J1} in his response to the EPR paper, is a remarkable feature of nonlocality in quantum theory, wherein one party (Alice) can remotely steer another distant party (Bob) by her choice of measurements.
EPR steering can be viewed as a quantum correlation (quantum resource) between quantum entanglement and Bell nonlocality, since it requires quantum entanglement as a fundamental resource for steering remote states, while EPR steering is not always sufficient to violate Bell inequality. Research on the relationship between Bell nonlocality, EPR steering, and quantum entanglement has made some progress \cite{J14,J15,J16}, but it remains an open question.
Unlike Bell nonlocality and quantum entanglement, EPR steering has a unique asymmetry, that is, one party can steer the another but not vice versa.
Because of its asymmetric properties, EPR steering has potential applications in  quantum secret sharing \cite{J2,J3,J4,J5}, quantum networks \cite{J6}, and quantum key distribution \cite{J7,J8}.

String theory is a promising candidate for a unified theory between general relativity and quantum mechanics.
Unlike general relativity, the string theory  predicted that the existence of the dilaton fields changes the properties of black holes \cite{J9,J10,J11,J12}.
The dilaton black holes are formed by gravitational systems coupled to Maxwell and dilaton fields, i.e., the Garfinkle-Horowitz-Strominger (GHS) dilaton black hole \cite{J10,J11}.
The Hawking effect \cite{J13} of the GHS black hole relies not only on the mass of the black hole, but also on its dilaton field, since the latter is also the source of gravity.
The Hawking effect for the dilaton black hole on quantum correlation, quantum coherence, and entropic uncertainty relation has been widely studied \cite{J44,J45,J46,J47,J48,J49,J50,J51,J52,J53,J54,J55,J56}.
However, the relationship between quantum correlations is still unclear in the background of the GHS dilaton black hole. Therefore,  studying the relationship between quantum steering, Bell nonlocality, and quantum entanglement in dilaton spacetime is one of the motivations for our work.

Another motivation for our work is to investigate the redistribution of the fermionic steering and the difference between the bosonic and fermionic steering in the background of the dilaton black hole.
Based on these motivations, it is assumed that our model involves three fermionic modes:
the first mode $A$ is observed by Alice at the asymptotically flat region;
the second mode $B$ is observed by Bob who hovers near the event horizon of the dilaton black hole; the third mode $\bar{B}$ observed by Anti-Bob is restricted by the event horizon.
By calculating the fermionic steering in dilaton spacetime,  we find that the $A \rightarrow B$ fermionic steerability is always larger than the $B \rightarrow A$ fermionic steerability, but the $A \rightarrow B$ bosonic steerability is always smaller than the $B \rightarrow A$ bosonic steerability \cite{J49}.
Furthermore, the fermionic steering can always survive, while the bosonic steering suffers a sudden death in curved spacetime.
We also find that the dilaton gravity can redistribute the fermionic steering, but cannot redistribute  Bell nonlocality, which indicates that the physically inaccessible steering is not nonlocal.
Finally, we obtain the relationships between the fermionic steering and entanglement in dilaton spacetime.
Therefore, we can understand another type of quantum correlation through one type of quantum correlation in dilaton spacetime.

Our paper is organized as follows.
In Sect. \ref{GSCDGE 1}, we introduce the quantification of quantum steering and the Clauser-Horne-Shimony-Holt (CHSH)  inequality.
In Sect. \ref{GSCDGE 2}, we discuss the quantization of the Dirac field in the background of the GHS dilaton black hole.
In Sect. \ref{GSCDGE 3}, we study the redistribution of the fermionic steering in dilaton spacetime.
In Sect. \ref{GSCDGE 4}, we obtain the monogamy relations between the fermionic  steering and entanglement in dilaton spacetime.
Finally, the Sect. \ref{GSCDGE 5} is devoted to the conclusion.

\section{  Quantification of quantum steering and CHSH inequality  \label{GSCDGE 1}}

Bell nonlocality, quantum steering, and quantum entanglement have a strict hierarchy for the mixed states. Interestingly,
Bell nonlocality can be indirectly detected by the notion of quantum steering \cite{J14}, and quantum steering can be indirectly detected by the concept of quantum entanglement \cite{J15,J16}.
We consider the density matrix of the X-state $\rho_{x}$ as
\begin{eqnarray}\label{R1}
\rho_{x} = \left(\!\!\begin{array}{cccccccc}
    \rho_{11} & 0 & 0 & \rho_{14} \\
    0 & \rho_{22} & \rho_{23} & 0 \\
    0 & \rho_{32} & \rho_{33} & 0 \\
    \rho_{41} & 0 & 0 & \rho_{44} \\
    \end{array}\!\!\right),
\end{eqnarray}
where $\rho_{ij}$ is the real element satisfying $\rho_{ij} =\rho_{ji}$.
As we all know, quantum entanglement of the bipartite states can be effectively identified by the concurrence.
The concurrence of the X-state $\rho_x$ given by Eq.(\ref{R1}) can be specifically shown as \cite{J17}
\begin{eqnarray}\label{R2}
C(\rho_x) = 2\max \left\{0, \left\lvert \rho_{14} \right\rvert - \sqrt{\rho_{22} \rho_{33}} , \left\lvert \rho_{23} \right\rvert - \sqrt{\rho_{11} \rho_{44}} \right\} .
\end{eqnarray}
For a general bipartite state $\rho_{AB}$ shared by Alice and Bob, the steering from Bob to Alice can be witnessed if the density matrix $\tau_{AB}$ defined as
\begin{eqnarray}\label{R3}
\tau_{AB} = \frac{\rho_{AB}}{\sqrt{3}} + \frac{3-\sqrt{3}}{3}\left(\rho_{A} \otimes \frac{I}{2} \right),
\end{eqnarray}
is entangled, where $\rho_A = \rm{Tr}_B(\rho_{AB})$ and $I$ is the two-dimension identity matrix \cite{J16,ZJ17}. Similarly, we can witness the steering from Alice to Bob when the density matrix $\tau_{BA}$ defined as
\begin{eqnarray}\label{R4}
\tau_{BA} = \frac{\rho_{AB}}{\sqrt{3}} + \frac{3-\sqrt{3}}{3}\left(\frac{I}{2} \otimes \rho_{B} \right),
\end{eqnarray}
is entangled, where $\rho_B = \rm{Tr}_A (\rho_{AB})$.

Through simple calculations, the matrix $\tau_{AB}$ of the X-state $\rho_x$ can be expressed as
\begin{eqnarray}\label{R5}
    \tau^{x}_{AB} = \left(\!\!\begin{array}{cccccccc}
        \frac{\sqrt{3}}{3} \rho_{11} + a & 0 & 0 & \frac{\sqrt{3}}{3} \rho_{14} \\
        0 & \frac{\sqrt{3}}{3} \rho_{22} + a & \frac{\sqrt{3}}{3} \rho_{23} & 0 \\
        0 & \frac{\sqrt{3}}{3} \rho_{32} & \frac{\sqrt{3}}{3} \rho_{33} + s & 0 \\
        \frac{\sqrt{3}}{3} \rho_{41} & 0 & 0 & \frac{\sqrt{3}}{3} \rho_{44} + s \\
        \end{array}\!\!\right),
\end{eqnarray}
with $a = \frac{3 - \sqrt{3}}{6} (\rho_{11} + \rho_{22})$ and $s = \frac{3 - \sqrt{3}}{6} (\rho_{33} + \rho_{44})$.
Using Eq.(\ref{R2}), the state $\tau^{x}_{AB}$ is entangled, if the state $\tau_{AB}$ satisfies inequality
\begin{eqnarray}\label{R6}
\left\lvert \rho_{14} \right\rvert^2 > L_a - L_b,
\end{eqnarray}
or
\begin{eqnarray}\label{R7}
\left\lvert \rho_{23} \right\rvert^2 > L_c - L_b,
\end{eqnarray}
where
\begin{eqnarray}\label{R8}
&L_a = \frac{2 - \sqrt{3}}{2} \rho_{11} \rho_{44} + \frac{2 + \sqrt{3}}{2} \rho_{22} \rho_{33} + \frac{1}{4} \left( \rho_{11} + \rho_{44} \right) \left( \rho_{22} + \rho_{33} \right),\notag\\
&L_b = \frac{1}{4} \left( \rho_{11} - \rho_{44} \right) \left( \rho_{22} - \rho_{33} \right),\\
&L_c = \frac{2 + \sqrt{3}}{2} \rho_{11} \rho_{44} + \frac{2 - \sqrt{3}}{2} \rho_{22} \rho_{33} + \frac{1}{4} \left( \rho_{11} + \rho_{44} \right) \left( \rho_{22} + \rho_{33} \right).\notag
\end{eqnarray}
Using a similar method, we find that the steering from Alice to Bob can be witnessed via one of the inequality
\begin{eqnarray}\label{R9}
\left\lvert \rho_{14} \right\rvert^2 > L_a + L_b,
\end{eqnarray}
or
\begin{eqnarray}\label{R10}
\left\lvert \rho_{23} \right\rvert^2 > L_c + L_b.
\end{eqnarray}
According to the inequality, the steerability from Bob to Alice  $S^{B \rightarrow A}$ is found to be
\begin{eqnarray}\label{R11}
S^{B \rightarrow A} = \max \left\{0 , \frac{8}{\sqrt{3}} [\left\lvert \rho_{14} \right\rvert^2 - L_a + L_b ] , \frac{8}{\sqrt{3}} [\left\lvert \rho_{23} \right\rvert^2 - L_c + L_b ] \right\}.
\end{eqnarray}
By exchanging the mode $A$ and the mode $B$, we can obtain  the steerability $S^{A \rightarrow B}$ as
\begin{eqnarray}\label{R12}
S^{A \rightarrow B} = \max \left\{0 , \frac{8}{\sqrt{3}} [\left\lvert \rho_{14} \right\rvert^2 - L_a - L_b ] , \frac{8}{\sqrt{3}} [\left\lvert \rho_{23} \right\rvert^2 - L_c - L_b ] \right\} .
\end{eqnarray}
The factor $\frac{8}{\sqrt{3}}$  guarantees that the steerability of the maximally entangled state is $1$.

The Bell inequality  can be violated in  quantum mechanics, which means that quantum mechanics cannot be redefined as a local realist theory.
As we all know, the typical Bell inequality is the CHSH inequality.
To study the relationship between quantum steering and Bell nonlocality, we use the CHSH inequality to verify the nonlocality of quantum steering.
The Bell operator for the CHSH inequality can be defined as
\begin{eqnarray}\label{H1}
\mathcal{B} = \mathbf{a} \cdot \sigma \otimes (\mathbf{b + b'}) \cdot \sigma + \mathbf{a'} \cdot \sigma \otimes (\mathbf{b - b'}) \cdot \sigma,
\end{eqnarray}
where $\mathbf{a}, \mathbf{a'}, \mathbf{b}$, and $\mathbf{b'}$ are unit vectors in $\mathbb{R}^3$, and $\sigma =(\sigma_1, \sigma_2, \sigma_3)$ denotes the vector of Pauli matrices.
To detect the Bell nonlocality of a state $\rho$, we employ the CHSH inequality and its expression of inequality for the two qubits to test local-realistic theories reads
\begin{eqnarray}\label{H2}
B(\rho ) =\left\lvert \rm{Tr} (\rho \mathcal{B} )\right\rvert \leqslant 2.
\end{eqnarray}
Thus, the requirement to violate the CHSH inequality is $B(\rho )> 2$, and the violation of this inequality implies the nonlocality of the state.
We need to find the maximal Bell signal $B(\rho )$, which for two-qubit systems can be equivalent to
\begin{eqnarray}\label{H3}
B(\rho ) = 2 \sqrt{\max_{i < j} (K_i + K_j)} ,
\end{eqnarray}
where $K_i$ and $K_j$ are the eigenvalues of the real symmetric matrix $K(\rho) =T^T_\rho T_\rho$, and $T = (t_{i j})$ represents the correlation matrix with $t_{i j} = \rm{Tr}[\rho \sigma_i \sigma_j]$.
For the two-qubit X-state, $K_1$, $K_2$, and $K_3$ take the specific form
\begin{eqnarray}\label{H4}
&K_1 = 4 (\left\lvert \rho_{14} \right\rvert + \left\lvert \rho_{23} \right\rvert )^2 ,\notag\\
&K_2 = 4 (\left\lvert \rho_{14} \right\rvert - \left\lvert \rho_{23} \right\rvert )^2 ,\notag\\
&K_3 = (\left\lvert \rho_{11} \right\rvert - \left\lvert \rho_{22} \right\rvert - \left\lvert \rho_{33} \right\rvert + \left\lvert \rho_{44} \right\rvert)^2.
\end{eqnarray}
As $K_1$ is greater than $K_2$, we can represent the maximal Bell signal as
\begin{eqnarray}\label{H5}
B(\rho_{x} ) = \max \left\{B_1, B_2\right\} ,
\end{eqnarray}
with $B_1 = 2 \sqrt{K_1 + K_2}$ and $B_2 = 2 \sqrt{K_1 + K_3}$ \cite{J18,J19}.
Note that the maximal  violation of the CHSH inequality for certain states is $2\sqrt{2} $,  and this bound can only be obtained by the maximally steered states.
We will use Eqs.(\ref{H2}), (\ref{H3}), and (\ref{H5}) to judge whether quantum steering is nonlocal in the dilaton black hole.

\section{Quantization of Dirac field in dilaton black hole\label{GSCDGE 2}}
Let us now introduce the massless Dirac equation in a general background spacetime \cite{J20,J21}
\begin{eqnarray}\label{R13}
[\gamma^a e^\mu_a(\partial_\mu + \Gamma_\mu)]\Psi = 0,
\end{eqnarray}
where $\gamma^a$ denotes the Dirac matrices, $\Gamma_\mu = \frac{1}{8}[\gamma^a,\gamma ^b]e^\nu_a e_{b\nu;\mu}$ is the spin connection coefficient, and the four-vectors $e^\mu_a$ represents the inverse of the tetrad $e^a_\mu$.
Note that the $e^a_\mu$ is defined by $g_{\mu\nu} = \eta_{ab} e^a_\mu e^b_\nu$ with $\eta_{ab} = \rm{diag}(-1,1,1,1)$.

The thermal Fermi-Dirac distribution of particles of the GHS dilaton black hole  with the Hawking temperature $T = \frac{1}{8\pi (M -D)}$ has been computed \cite{J22,J23,J24}.
It is well known that the presence of such radiation is called the Hawking effect.
The metric for the GHS  dilaton black hole reads \cite{J25,J27}
\begin{eqnarray}\label{R14}
ds^2 = -\bigg(\frac{r-2M}{r-2D}\bigg)dt^2 + \bigg(\frac{r-2M}{r-2D}\bigg)^{-1} dr^2 + r(r-2D)d\Omega^2,
\end{eqnarray}
where $M$ and $D$ represent the mass and the dilaton of the black hole, respectively.
Throughout this paper, we set $\hbar = G = c = \kappa_B = 1$ for convenience.
In addition, the dilaton $D$ and the mass $M$ should satisfy the relationship $D < M$.
To separate the Dirac equation in the following discussion, we utilize a tetrad as
\begin{eqnarray}\label{R15}
e^a_\mu = \rm{diag}\bigg(\sqrt{f},\frac{1}{\sqrt{f}},\sqrt{r\tilde{r}},\sqrt{r\tilde{r}} \sin \theta\bigg),
\end{eqnarray}
where $f=\frac{(r-2M)}{\tilde{r}}$ and $\tilde{r}=r-2D$.
According to Eq.(\ref{R13}), the massless Dirac equation in the GHS dilaton black hole can be specifically represented as
\begin{eqnarray}\label{R16}
&&-\frac{r_0}{\sqrt{f}} \frac{\partial \Psi}{\partial t} + \sqrt{f} \gamma_1\bigg(\frac{\partial}{\partial r} + \frac{r-D}{r \tilde{r}} + \frac{1}{4f} \frac{df}{dr}\bigg)\Psi\nonumber\\
 &&+\frac{\gamma_2}{\sqrt{r\tilde{r}}}\bigg(\frac{\partial}{\partial \theta}+\frac{\cot \theta}{2}\bigg)\Psi + \frac{\gamma_3}{\sqrt{r\tilde{r}}\sin \theta} \frac{\partial \Psi}{\partial \varphi} = 0.
\end{eqnarray}
If we use $\Psi = f^{-\frac{1}{4}} \Phi$ \cite{J28}, we can solve the Dirac equation near the event horizon of the black hole.
For  the exterior region and  interior region of the event horizon, we obtain the following positive frequency outgoing solutions \cite{J29,J30,J31}
\begin{eqnarray}\label{R17}
\Psi^+_{out,\boldmath{k}} = \mathcal{J} e^{-i \omega \mathcal{O}},
\end{eqnarray}
\begin{eqnarray}\label{R18}
\Psi^+_{in,\boldmath{k}} = \mathcal{J} e^{i \omega \mathcal{O}},
\end{eqnarray}
where $\mathcal{O} = t - r_*$, $\mathcal{J}$ is a four-component Dirac spinor, and $\boldmath{k}$ is the wave vector that can be used to label the modes.
Using Eq.(\ref{R17}) and Eq.(\ref{R18}),  the Dirac field can be expanded as
\begin{eqnarray}\label{R19}
\Psi = \sum_{\sigma} \int d \boldmath{k}[\hat{a}^{\sigma}_{\boldmath{k}} \Psi^+_{\sigma, \boldmath{k}} + \hat{b}^{\sigma \dagger}_{\boldmath{k}} \Psi^-_{\sigma, \boldmath{k}}],
\end{eqnarray}
where $\sigma = (in, out)$, $\hat{a}^{\sigma}_{\boldmath{k}}$ and $\hat{b}^{\sigma \dagger}_{\boldmath{k}}$ are the fermion annihilation and antifermion creation operators acting on the quantum state, respectively.
The annihilation operator and creation operator satisfy the canonical anticommutation relations
$\left\{\hat{a}^{out}_{\boldmath{k}},\hat{a}^{out \dagger}_{\boldmath{k'}}\right\} = \left\{\hat{a}^{in}_{\boldmath{k}},\hat{a}^{in \dagger }_{\boldmath{k'}}\right\} = \left\{\hat{b}^{out}_{\boldmath{k}},\hat{b}^{out \dagger}_{\boldmath{k'}}\right\} = \left\{\hat{b}^{in}_{\boldmath{k}},\hat{b}^{in \dagger }_{\boldmath{k'}}\right\} = \delta_{\boldmath{k} \boldmath{k'}}$. The  annihilation operators $\hat{a}^{\sigma}_{\boldmath{k}}$ define the dilaton vacuum as  $\hat{a}^{out}_{\boldmath{k}} \vert 0 \rangle_D = \hat{a}^{in}_{\boldmath{k}} \vert 0 \rangle_D = 0$. Usually, we call the dilaton  modes $\Psi^{\pm}_{\sigma, \boldmath{k}}$.

The complete basis for positive energy modes, i.e., the Kruskal modes introduced by Domour-Ruffini \cite{J17}, can make analytic continuations of Eqs.(\ref{R17}) and (\ref{R18}).
We can also use the Kruskal modes to expand the Dirac field
\begin{eqnarray}\label{R20}
\Psi = \sum_{\sigma} \int d \boldmath{k} \frac{1}{\sqrt{2 \cosh (4 \pi (M - D)\omega)}} \left[ \hat{c}^{\sigma}_{\boldmath{k}} \Phi^+_{\sigma, \boldmath{k}} + \hat{d}^{\sigma \dagger}_{\boldmath{k}} \Phi^-_{\sigma, \boldmath{k}}\right],
\end{eqnarray}
where $\hat{c}^{\sigma}_{\boldmath{k}}$ and $\hat{d}^{\sigma \dagger}_{\boldmath{k}}$ are the fermion annihilation and antifermion creation operators acting on the Kruskal vacuum, respectively.
It can be seen from Eq.(\ref{R19}) and Eq.(\ref{R20}) that the Dirac field is decomposed into the Kruskal and dilaton modes, respectively.
Thus, we can obtain the Bogoliubov transformations between Kruskal and dilaton modes.
Using the Bogoliubov transformations, we can get the relations between the Kruskal and dilaton operators that take the forms
\begin{eqnarray}\label{R21}
\hat{c}^{out}_{\boldmath{k}} = \frac{1}{\sqrt{e^{- 8 \pi (M - D) \omega } + 1} } \hat{a}^{out}_{\boldmath{k}} - \frac{1}{\sqrt{e^{8 \pi (M - D) \omega } + 1} } \hat{b}^{out \dagger}_{\boldmath{k}},\notag\\
\hat{c}^{out \dagger}_{\boldmath{k}} = \frac{1}{\sqrt{e^{- 8 \pi (M - D) \omega } + 1} } \hat{a}^{out \dagger}_{\boldmath{k}} - \frac{1}{\sqrt{e^{8 \pi (M - D) \omega } + 1} } \hat{b}^{out}_{\boldmath{k}}.
\end{eqnarray}
Since the GHS dilaton black hole can be divided into physically accessible and inaccessible regions, the  ground and only excited states in Kruskal spacetime become the two-mode squeezed state in the dilaton black hole.
After properly normalizing the state vector, the Kruskal vacuum and only excited states in  dilaton spacetime read
\begin{eqnarray}\label{R22}
\nonumber&\vert 0 \rangle_K = \frac{1}{\sqrt{e^{- 8 \pi (M - D) \omega } + 1} } \vert 0 \rangle_{out} \vert 0 \rangle_{in} + \frac{1}{\sqrt{e^{8 \pi (M - D) \omega } + 1} } \vert 1 \rangle_{out} \vert 1 \rangle_{in},\\
&\vert 1 \rangle_K = \vert 1 \rangle_{out} \vert 0 \rangle_{in},
\end{eqnarray}
where $\left\{\vert n \rangle_{out}\right\} $ and $\left\{\vert n \rangle_{in}\right\} $
correspond to the orthonormal bases for the outside and inside regions of the event horizon, respectively.

\section{ Redistribution of fermionic steering in dilaton spacetime \label{GSCDGE 3}}
We initially  assume that Alice and Bob share a maximally entangled state in the asymptotically flat region of the dilaton black hole, which can be written in the following form
\begin{eqnarray}\label{R23}
\vert \phi_{AB} \rangle  = \frac{1}{\sqrt{2}}(\vert 0 \rangle_A \vert 0 \rangle_B + \vert 1 \rangle_A \vert 1 \rangle_B ),
\end{eqnarray}
where the modes $A$ and $B$ are observed by Alice and Bob, respectively.
Then, Bob hovers near the event horizon of the dilaton black hole and Alice continues to stay at the asymptotically flat region. Therefore, Bob can detect a thermal Fermi-Dirac distribution of particles, meaning that his detector is found to be excited.
We can use the dilaton modes of Eq.(\ref{R22}) to rewrite the initial entangled state
\begin{eqnarray}\label{R24}
\vert \phi_{AB \bar{B} } \rangle  =& \frac{1}{\sqrt{2}}\bigg(\frac{1}{\sqrt{e^{-8 \pi (M - D)\omega + 1 }}}\vert 0 \rangle_A \vert 0 \rangle_B \vert 0 \rangle_{\bar{B}} + \frac{1}{\sqrt{e^{8 \pi (M - D) \omega + 1 }}} \vert 0 \rangle_A \vert 1 \rangle_B \vert 1 \rangle_{\bar{B}} \\ \nonumber
 &+ \vert 1 \rangle_A \vert 1 \rangle_B \vert 0 \rangle_{\bar{B}} \bigg).
\end{eqnarray}
Here, the physically inaccessible mode $\bar{B}$ is observed by Anti-Bob inside the event horizon. The density matrix of quantum state $\phi_{AB \bar{B} }$ in the orthonormal basis $\left\{\vert 0,0,0 \rangle , \vert 0,0,1 \rangle , \vert 0,1,0 \rangle , \vert 0,1,1 \rangle , \vert 1,0,0 \rangle , \vert 1,0,1 \rangle , \vert 1,1,0 \rangle , \vert 1,1,1 \rangle\right\}$ can be expressed as
\begin{eqnarray}\label{R25}
    \rho _{AB \bar{B}} = \frac{1}{2} \left(\!\!\begin{array}{cccccccc}
        \frac{1}{e^{-8 \pi (M - D)\omega } + 1} & 0 & 0 & \frac{1}{\sqrt{e^{-8 \pi (M - D)\omega }+ e^{8 \pi (M - D) \omega } + 2}} & 0 & 0 & \frac{1}{\sqrt{e^{-8 \pi (M - D)\omega }+ 1}} & 0\\
        0 & 0 & 0 & 0 & 0 & 0 & 0 & 0\\
        0 & 0 & 0 & 0 & 0 & 0 & 0 & 0 \\
        \frac{1}{\sqrt{e^{-8 \pi (M - D)\omega }+ e^{8 \pi (M - D) \omega } + 2}} & 0 & 0 & \frac{1}{e^{8 \pi (M - D)\omega } + 1} & 0 & 0 & \frac{1}{\sqrt{e^{8 \pi (M - D) \omega }+ 1}} & 0\\
        0 & 0 & 0 & 0 & 0 & 0 & 0 & 0 \\
        0 & 0 & 0 & 0 & 0 & 0 & 0 & 0 \\
        \frac{1}{\sqrt{e^{-8 \pi (M - D)\omega }+ 1}} & 0 & 0 & \frac{1}{\sqrt{e^{8 \pi (M - D) \omega }+ 1}} & 0 & 0 & 1 & 0\\
        0 & 0 & 0 & 0 & 0 & 0 & 0 & 0 \\
        \end{array}\!\!\right).
\end{eqnarray}

\subsection{ Physically accessible steering \label{GSCDGE 31}}
Because the exterior region of the black hole is causally disconnected from the interior region, Alice and Bob cannot approach the mode $\bar{B}$.
The physically accessible information is encoded in the mode $A$ described by Alice and the mode $B$  described by Bob.
Taking the trace over the mode $\bar{B} $  inside the event horizon, we obtain a mixed density matrix of Alice and Bob
\begin{eqnarray}\label{R26}
    \rho _{AB} =\frac{1}{2}  \left(\!\!\begin{array}{cccccccc}
        \frac{1}{e^{-8 \pi (M - D)\omega } + 1} & 0 & 0 & \frac{1}{\sqrt{e^{-8 \pi (M - D)\omega }+ 1}} \\
        0 & \frac{1}{e^{8 \pi (M - D)\omega } + 1} & 0 & 0 \\
        0 & 0 & 0 & 0 \\
        \frac{1}{\sqrt{e^{-8 \pi (M - D)\omega }+ 1}} & 0 & 0 & 1 \\
        \end{array}\!\!\right),
\end{eqnarray}
in the basis $\left\{\vert 00 \rangle , \vert 01 \rangle , \vert 10 \rangle , \vert 11 \rangle\right\} $. Employing Eqs.(\ref{R11}) and (\ref{R12}), quantum steering from Alice to Bob and quantum steering from Bob to Alice  are found to be
\begin{eqnarray}\label{R27}
    S^{A \rightarrow B} = \max \left\{0, \frac{1}{e^{-8 \pi (M - D)\omega } + 1} - \frac{1}{\sqrt{3} }\frac{1}{e^{-8 \pi (M - D)\omega }+ e^{8 \pi (M - D) \omega } + 2}\right\} ,
\end{eqnarray}
\begin{eqnarray}\label{R28}
    S^{B \rightarrow A} = \max \left\{0, \frac{1}{e^{-8 \pi (M - D)\omega } + 1} - \frac{1}{\sqrt{3} }\frac{1}{e^{8 \pi (M - D)\omega } + 1}\right\}.
\end{eqnarray}
$S>0$ indicates the presence of quantum steering, and $S=0$ indicates the absence of quantum steering.
Using Eq.(\ref{H5}), the maximal Bell signal can be writen as
\begin{eqnarray}\label{R29}
    B(\rho_{AB} ) =  \frac{2}{\sqrt{e^{-8 \pi (M - D)\omega }+ 1}}\sqrt{1 + \frac{1}{e^{-8 \pi (M - D)\omega } + 1}}  .
\end{eqnarray}

\begin{figure}
\begin{minipage}[t]{0.5\linewidth}
\centering
\includegraphics[width=3.2in,height=5.6cm]{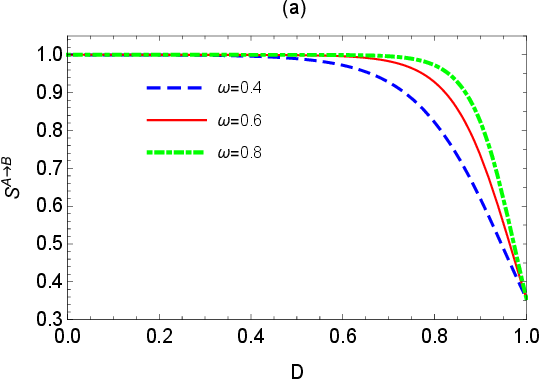}
\label{fig11}
\end{minipage}%
\begin{minipage}[t]{0.5\linewidth}
\centering
\includegraphics[width=3.2in,height=5.6cm]{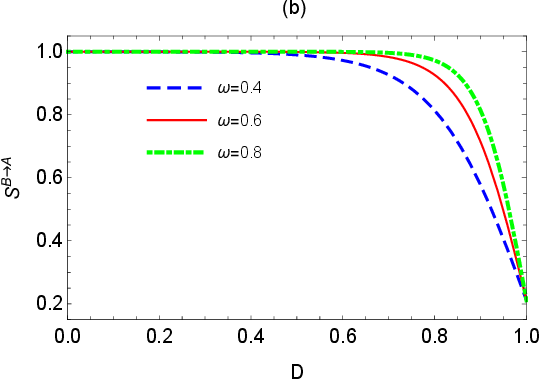}
\label{fig12}
\end{minipage}%

\begin{minipage}[t]{0.5\linewidth}
\centering
\includegraphics[width=3.2in,height=5.6cm]{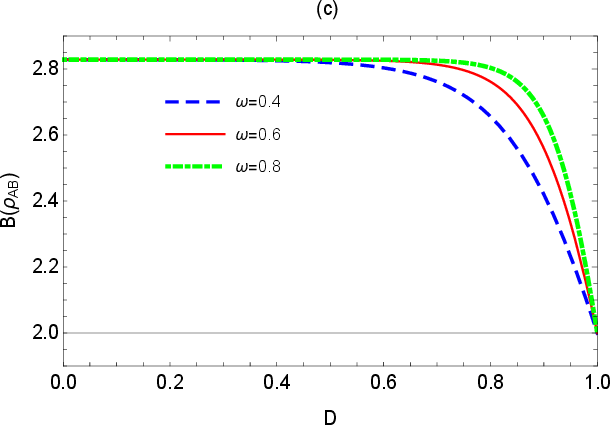}
\label{fig13}
\end{minipage}%
\begin{minipage}[t]{0.5\linewidth}
\centering
\includegraphics[width=3.5in,height=6.50cm]{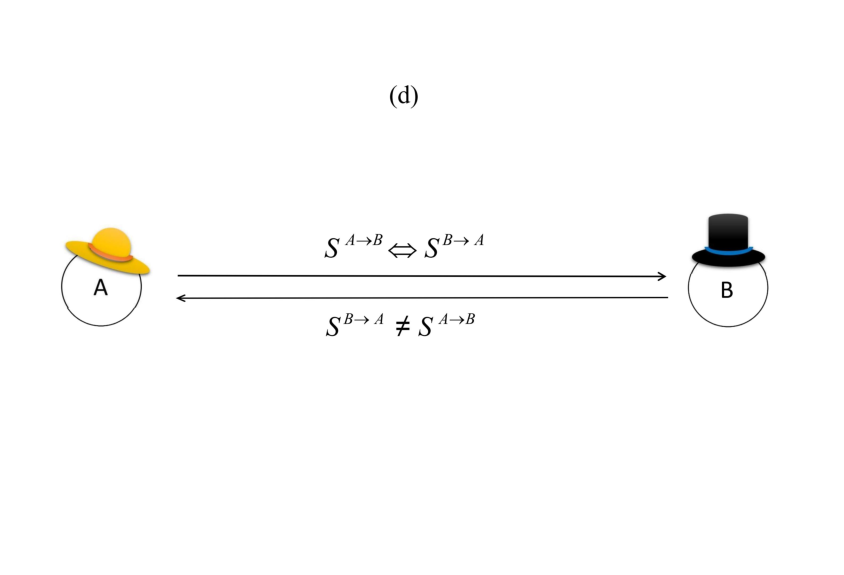}
\label{fig14}
\end{minipage}%

\caption{(a), (b), and (c) the fermionic steerability and the maximal Bell signal between Alice and Bob as a function of the dilaton $D$, for different $\omega $.
We set the parameter $M = 1$.
(d) an example of relationship of quantum steering between Alice and Bob.}
\label{Fig1}
\end{figure}

We plot the $A \rightarrow B$ steering, $B \rightarrow A$ steering, and the maximal Bell signal as a function of the dilation $D$ and plot the relationship of quantum steering between modes $A$ and $B$, as shown in Fig.\ref{Fig1}.
It can be seen from Fig.\ref{Fig1}(a) and (b) that the fermionic steering monotonically decreases to a fixed value with increasing dilaton $D$, while the bosonic steering  first irreversibly degenerates and then undergoes sudden death under the influence of the dilaton \cite{J49}.
We find that the fermionic steerability from Alice to Bob is always larger than the fermionic steerability from Bob to Alice, while  the bosonic  steerability from Alice to Bob is always smaller than the bosonic  steerability from Bob to Alice in dilaton spacetime.
If we use the steerability from Alice to Bob over the steerability from Bob to Alice for relativistic quantum information tasks, we should use the fermionic steering rather than the bosonic steering.
These results suggest that the fermionic steering contrasts sharply with the bosonic steering due to the difference between the Fermi-Dirac statistic and the Bose-Einstein statistic in dilaton spacetime.
From Fig.\ref{Fig1}(a) and (b), we can also see that the fermionic steering does not depend on the frequency  of the modes $\omega$  in the limit of an extremely dilaton black hole ($D \rightarrow M$).
In other words, the fermionic steering is not affected by the frequency in the limit of $D \rightarrow M$.
In Fig.\ref{Fig1}(c), we also find that $B(\rho_{AB} )$ is equal to 2 for $D \rightarrow M$, indicating that there is no Bell nonlocality in quantum state $\rho_{AB}$. This means that the fermionic steering cannot be considered to be nonlocal in the extreme dilaton black hole.
In Fig.\ref{Fig1}(d),  each circle represents an observer, and the arrows connecting two observers describe the bipartite steering relationship between them.
Based on the characteristics of steering, quantum steering in the bipartite system includes two-way steering, one-way steering, and no-way steering.
It is easy to  find  that the physically accessible steering between Alice and Bob is a two-way steering in curved spacetime.

\subsection{Physically inaccessible steering \label{GSCDGE 32}}
Next, we will explore the steering between Alice and Anti-Bob, and the steering between Bob and Anti-Bob under the influence of the dilaton.
Since Anti-Bob is inside the event horizon of the black hole, we refer to this type of quantum steering as a ``physically inaccessible steering''.

Firstly, we study the fermionic steering between Alice and Anti-Bob.
Tracing over the mode $B$ observed by Bob, we can get the density matrix between the modes $A$ and $\bar B$ as
\begin{eqnarray}\label{R30}
    \rho _{A\bar B} =\frac{1}{2}  \left(\!\!\begin{array}{cccccccc}
        \frac{1}{e^{-8 \pi (M - D)\omega } + 1} & 0 & 0 & 0 \\
        0 & \frac{1}{e^{8 \pi (M - D)\omega } + 1} & \frac{1}{\sqrt{e^{8 \pi (M - D) \omega }+ 1}} & 0 \\
        0 & \frac{1}{\sqrt{e^{8 \pi (M - D) \omega }+ 1}} & 1 & 0 \\
        0 & 0 & 0 & 0 \\
        \end{array}\!\!\right).
\end{eqnarray}
Using Eqs.(\ref{R11}), (\ref{R12}), and (\ref{H5}), the fermionic steering and the maximal Bell signal between Alice and Anti-Bob can be written as
\begin{eqnarray}\label{R31}
    S^{A \rightarrow \bar{B} } = \max \left\{0,\frac{1}{e^{8 \pi (M - D)\omega } + 1} \bigg[1 - \frac{1}{\sqrt{3} } \frac{1}{e^{-8 \pi (M - D)\omega } + 1}\bigg]\right\} ,
\end{eqnarray}
\begin{eqnarray}\label{R32}
    S^{\bar{B} \rightarrow A} = \max \left\{0,\frac{1}{e^{8 \pi (M - D)\omega } + 1} - \frac{1}{\sqrt{3} } \frac{1}{e^{-8 \pi (M - D)\omega } + 1}\right\} ,
\end{eqnarray}
and
\begin{eqnarray}\label{R33}
    B(\rho_{A\bar{B}} ) = \frac{2 }{\sqrt{e^{8 \pi (M - D)\omega }+ 1}}\sqrt{1 + \frac{1}{e^{8 \pi (M - D)\omega } + 1}} .
\end{eqnarray}

\begin{figure}
\begin{minipage}[t]{0.5\linewidth}
\centering
\includegraphics[width=3.2in,height=5.6cm]{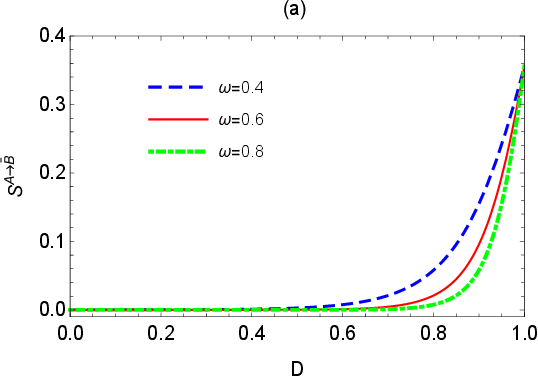}
\includegraphics[width=3.2in,height=5.6cm]{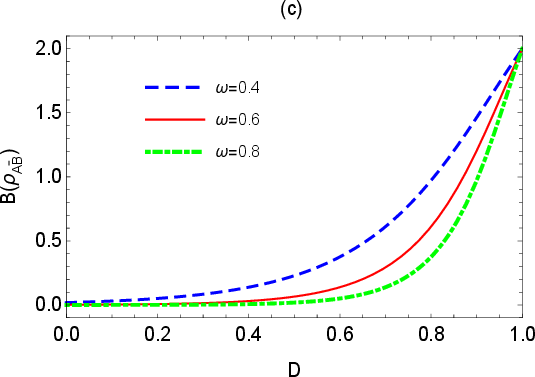}
\label{fig21}
\end{minipage}%
\begin{minipage}[t]{0.55\linewidth}
\centering
\includegraphics[width=3.2in,height=5.6cm]{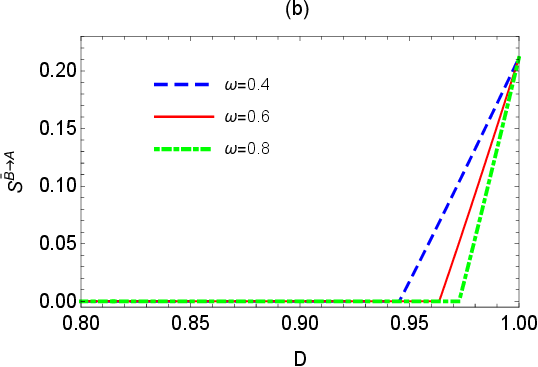}
\includegraphics[width=3.1in,height=4.5cm]{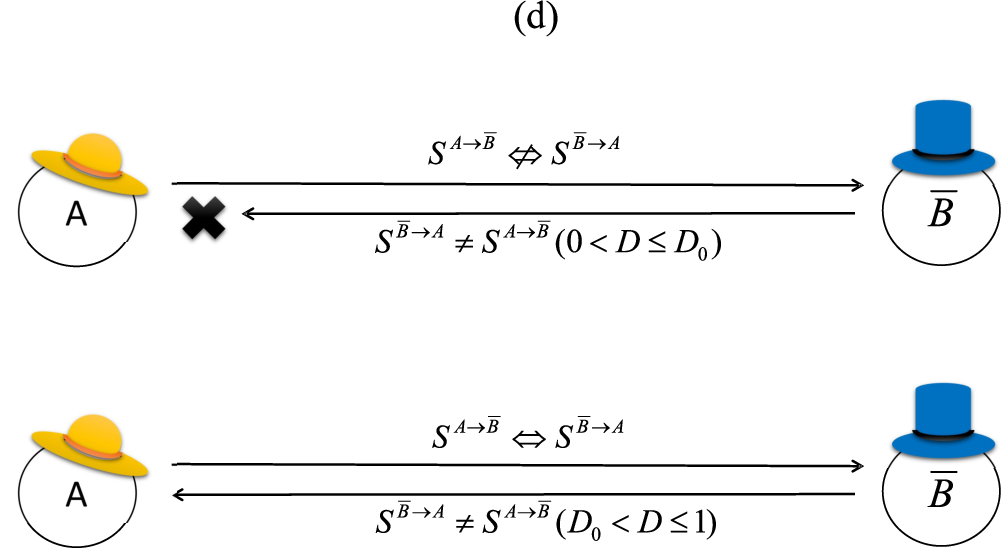}
\label{fig22}
\end{minipage}%

\caption{(a), (b), and (c) the fermionic steerability and the maximal Bell signal between Alice and Anti-Bob as a functon of the dilaton $D$, for different $\omega $.
We set the parameter  $M = 1$.
(d) an example of relationship of quantum steering between Alice and Anti-Bob.}
\label{Fig2}
\end{figure}

The influence of the dilaton $D$ of the black hole on the fermionic steerability and the maximal Bell signal between Alice and Anti-Bob for different frequencies $\omega$ are plotted in Fig.\ref{Fig2}(a), (b), and (c).
And the illustration in Fig.\ref{Fig2}(d) depicts the relationship of quantum steering between Alice and Anti-Bob. From Fig.\ref{Fig2}(a) and (b), we can see that the fermionic steering between Alice and Anti-Bob can be generated by gravitational effect, and the $\bar{B} \rightarrow A$ steering undergoes the sudden birth with the dilaton $D$.
In addition, the  $A \rightarrow \bar{B}$ steerability is always larger than the $\bar{B} \rightarrow A$   steerability in dilaton spacetime. We also see from Fig.\ref{Fig2}(a) and (b) that, for the given dilaton $D$, the steering between Alice and Anti-Bob decreases monotonically with the increase of the $\omega$.
Specially, for the non-dilaton and extreme dilaton black holes, quantum steering is independent of the frequency $\omega$.
In Fig.\ref{Fig2}(c), we find that $B(\rho_{AB} )$ is always less than  or equal to  2, which means that there is no Bell nonlocality between Alice and Anti-Bob.
Two different possibilities of bipartite steering between Alice and Anti-Bob are depicted in Fig.\ref{Fig2}(d).
According to the steering asymmetry of Alice and Anti-Bob, the fermionic steering exhibits unique directionality, which may lead to one-way steering, that is, Alice can steer Anti-Bob, but Anti-Bob cannot steer Alice. Note that the dilaton for the sudden birth of steering $S^{\bar{B} \rightarrow A}$ is $D_0 = M - \frac{1}{8\pi \omega} \ln\sqrt{3} $.
For $0<D \leq D_0$, we can obtain $S^{A\rightarrow \bar{B}} > 0$ and $S^{\bar{B} \rightarrow A} = 0$, which corresponds to one-way steering in Fig.\ref{Fig2}(d).
For $  D_0<D\leq1$, we find $S^{A\rightarrow \bar{B}} > 0$ and $S^{\bar{B} \rightarrow A} > 0$, so the steering between Alice and Anti-Bob is a two-way steering.

In addition, we also study the fermionic steering between Bob and Anti-Bob.
We trace over the mode $A$ and then get the density
matrix for the modes $B$ and $\bar B$
\begin{eqnarray}\label{R34}
    \rho _{B \bar B} =\frac{1}{2}  \left(\!\!\begin{array}{cccccccc}
        \frac{1}{e^{-8 \pi (M - D)\omega } + 1} & 0 & 0 & \frac{1}{\sqrt{e^{-8 \pi (M - D)\omega }+ e^{8 \pi (M - D) \omega } + 2}} \\
        0 & 0 & 0 & 0 \\
        0 & 0 & 1 & 0 \\
        \frac{1}{\sqrt{e^{-8 \pi (M - D)\omega }+ e^{8 \pi (M - D) \omega } + 2}} & 0 & 0 & \frac{1}{e^{8 \pi (M - D)\omega } + 1} \\
        \end{array}\!\!\right).
\end{eqnarray}
The fermionic steering and the maximal Bell signal between Bob and Anti-Bob can be expressed as
\begin{eqnarray}\label{R35}
    S^{B \rightarrow \bar{B} } = \max \left\{0,\frac{1}{e^{8 \pi (M - D)\omega } + 1}\bigg[\frac{1}{e^{-8 \pi (M - D)\omega } + 1} - \frac{1}{\sqrt{3} } \bigg]\right\} ,
\end{eqnarray}
\begin{eqnarray}\label{R36}
    S^{\bar{B} \rightarrow B} = \max \left\{0,\frac{1}{e^{-8 \pi (M - D)\omega } + 1}\bigg[\frac{1}{e^{8 \pi (M - D)\omega } + 1} - \frac{1}{\sqrt{3} } \bigg]\right\} ,
\end{eqnarray}
and
\begin{eqnarray}\label{R37}
    B(\rho_{B\bar{B}} ) = \frac{2 }{\sqrt{e^{-8 \pi (M - D)\omega }+ e^{8 \pi (M - D) \omega } + 2}} .
\end{eqnarray}

\begin{figure}
\begin{minipage}[t]{0.5\linewidth}
\centering
\includegraphics[width=3.2in,height=5.6cm]{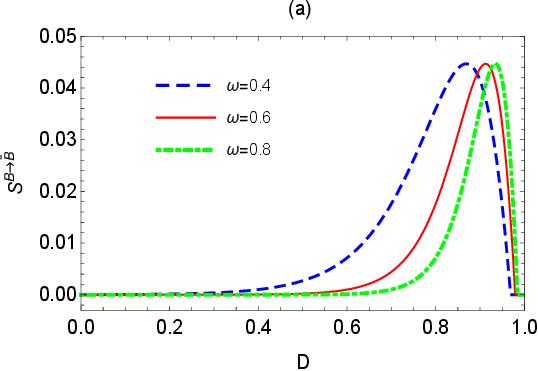}
\includegraphics[width=3.2in,height=5.6cm]{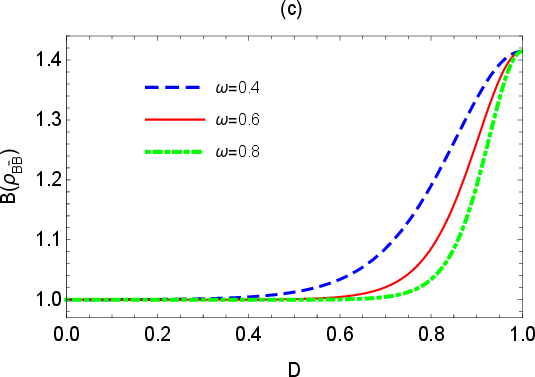}
\label{fig31}
\end{minipage}%
\begin{minipage}[t]{0.5\linewidth}
\centering
\includegraphics[width=3.2in,height=5.6cm]{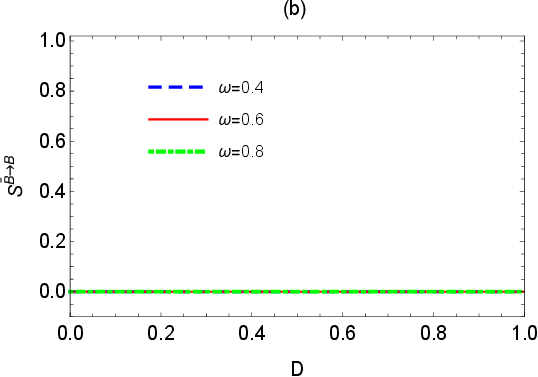}
\includegraphics[width=3.0in,height=4.5cm]{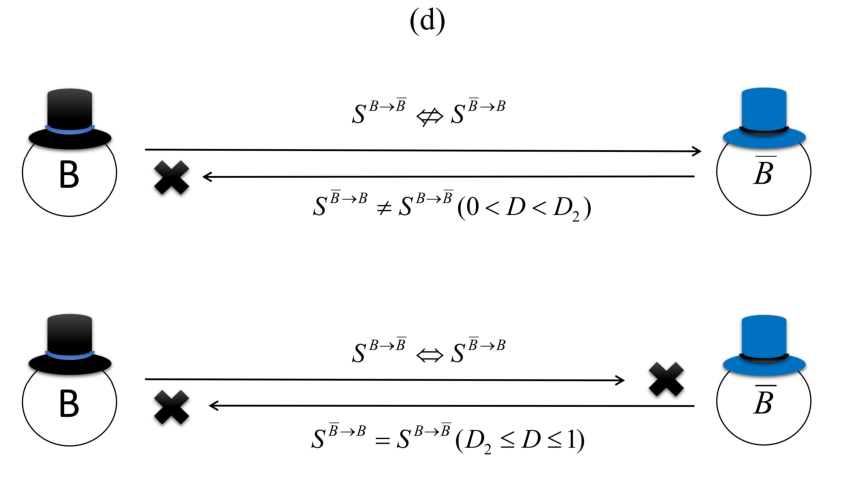}
\label{fig32}
\end{minipage}%

\caption{(a), (b), and (c) the fermionic steerability and the maximal Bell signal between Bob and Anti-Bob as a functon of the dilation $D$, for different $\omega $.
We set the parameter  $M = 1$.
(d) an example of relationship of quantum steering between Bob and Anti-Bob.}
\label{Fig3}
\end{figure}

In Fig.\ref{Fig3},  we plot quantum steering and the maximal Bell signal between Bob and Anti-Bob as a function of the dilaton $D$, as well as its relationship in dilaton spacetime.
From Fig.\ref{Fig3}(a), we can see that the fermionic steering $S^{B \rightarrow \overline{B}}$ first increases from zero to the maximum and then suffers sudden death with the growth of the dilaton $D$.
In Fig.\ref{Fig3}(b), we can find that the fermionic steering $S^{\overline{B} \rightarrow B}$ is always equal to zero. The dilaton for the maximal steering is $D_1 = M - \frac{1}{8\pi \omega }[\ln(\sqrt{3} + 1) - \ln(\sqrt{3} - 1)]$, and the dilaton for the sudden death of quantum steering is $D_2 = M + \frac{1}{8\pi \omega } \ln(\sqrt{3} - 1) $.
In Fig.\ref{Fig3}(c), we can also find that the maximal Bell signal $B(\rho_{AB} )$ is always less than $2$, indicating that there  is no Bell nonlocality between Bob and Anti-Bob.
Fig.\ref{Fig3}(d) shows two relationships of quantum steering  between Bob and Anti-Bob. The conditions of its relationship are given as follows: (i)
the condition $0<D < D_2$ means one-way steering $S^{B\rightarrow \bar{B}} > 0$ and $S^{\bar{B} \rightarrow B} = 0$; (ii)
the condition  $D_2\leq D\leq1$ implies no-way steering $S^{B\rightarrow \bar{B}} = 0$ and $S^{\bar{B} \rightarrow B} = 0$.

We compare Figs.\ref{Fig1}-\ref{Fig3} in order to relate the two fundamental physical phenomena of quantum steering and Bell nonlocality.
We conclude that the fermionic steering cannot be considered nonlocality in the extreme dilaton black hole.
From Figs.\ref{Fig2}-\ref{Fig3}, we can find that the physically inaccessible steering is also not nonlocal.
We can also find that quantum steering and the maximal Bell signal are not affected by the frequency of the fermionic field for the extreme dilaton black hole.
In other words, quantum steering and the maximal Bell signal for different frequencies approach the same values in the extreme dilaton black hole.

\subsection{Asymmetry of quantum steering \label{GSCDGE 33}}
Unlike quantum entanglement and Bell nonlocality, quantum steering has asymmetric. Therefore, we distinguish quantum steering into three cases: (i) the first one corresponds to no-way steering, showing that the state is nonsteerable in any direction; (ii) the second case is two-way steering, that is to see, the state can be steerable in both directions;
(iii) the third case is one-way steering, indicating that the state is steerable only in one direction.
The last case reflects the asymmetric nature of quantum steering.
To measure the degrees of asymmetry in the dilaton black hole, we define the steering asymmetries  between the modes $A$ and $B$, $A$ and $\bar{B}$, or $B$ and $\bar{B}$ as
\begin{eqnarray}\label{R38}
    S^{\Delta}_{AB} = \left\lvert S^{A \rightarrow B} - S^{B \rightarrow A} \right\rvert ,
\end{eqnarray}
\begin{eqnarray}\label{R39}
    S^{\Delta}_{A \bar{B} } = \left\lvert S^{A \rightarrow  \bar{B}} - S^{ \bar{B} \rightarrow A} \right\rvert ,
\end{eqnarray}
\begin{eqnarray}\label{R40}
    S^{\Delta}_{B \bar{B} } = \left\lvert S^{B \rightarrow  \bar{B}} - S^{ \bar{B} \rightarrow B} \right\rvert .
\end{eqnarray}

\begin{figure}
\begin{minipage}[t]{0.33\linewidth}
\centering
\includegraphics[width=2.1in,height=4.1cm]{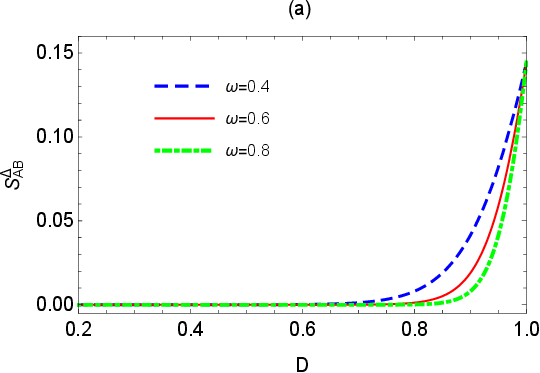}
\label{fig41}
\end{minipage}%
\begin{minipage}[t]{0.33\linewidth}
\centering
\includegraphics[width=2.1in,height=4.1cm]{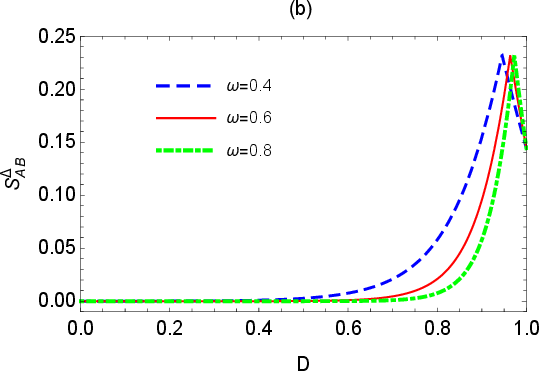}
\label{fig42}
\end{minipage}%
\begin{minipage}[t]{0.33\linewidth}
\centering
\includegraphics[width=2.1in,height=4.1cm]{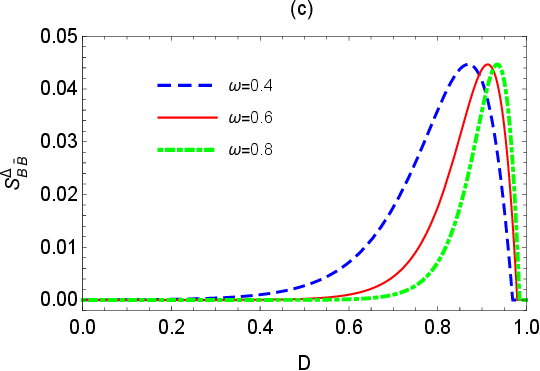}
\label{fig43}
\end{minipage}%

\caption{The steering asymmetry $S^{\Delta}_{AB}$, $S^{\Delta}_{A \bar{B}}$, and $S^{\Delta}_{B \bar{B}}$ as a function of the dilaton $D$, for different $\omega$.
We set the parameter  $M = 1$. }
\label{Fig4}
\end{figure}

Fig.\ref{Fig4} shows how the dilaton $D$  of the black hole
influences the steering asymmetry for different frequencies $\omega$.
It can be seen that, as the dilaton $D$ increases, the steering asymmetry $S^{\Delta}_{AB}$ increases, while the steering asymmetry $S^{\Delta}_{A \bar{B} }$ increases from zero to the maximum and then decreases to a fixed value, and the steering asymmetry $S^{\Delta}_{B \bar{B} }$ suffers sudden death.
In Fig.\ref{Fig4}(a), we find that the steering asymmetry between Alice and Bob has the case of two-way steering, which means that they can steer each other.
Fig.\ref{Fig4}(c) presents quantum steering in one direction (one-way steering), indicating that quantum steering between $B$ and $\bar{B}$ is completely asymmetric.
From Fig.\ref{Fig4}(b), we can also find that  maximal steering
asymmetry between $A$ and $\bar{B}$ shows the the transformation between one-way steering and two-way steering. In other words, for the point $D = D_0$, the system $ \rho _{A\bar B}$ is experiencing a transformation from one-way steering to two-way steering in dilaton spacetime.
In addition, the steering asymmetry is also independent of the frequency $\omega $ for the non-dilaton and the extreme dilaton black hole.

\section{Monogamy relations between quantum steering and entanglement in dilaton spacetime \label{GSCDGE 4}}

In the previous section, we have studied the relationship between quantum steering and Bell nonlocality in detail, but the relationship between steering and entanglement is still not very clear  in dilaton spacetime.
As is well known, quantum steering is an intermediate form of quantum correlation between Bell nonlocality and quantum entanglement and a good connection between Bell nonlocality and quantum entanglement \cite{J43}.
Therefore, we try to find the relationships between the fermionic steering and entanglement in dilaton spacetime.

According to Eq.(\ref{R2}), we calculate the concurrence $C(\rho _{AB})$ between Alice and Bob, $C(\rho _{A\bar{B} })$ between Alice and Anti-Bob, as well as $C(\rho _{B\bar{B} })$ between Bob and Anti-Bob, which can be written as
\begin{eqnarray}\label{R41}
    C(\rho _{AB}) = \frac{1}{\sqrt{e^{-8 \pi (M - D)\omega} +1}  } ,
\end{eqnarray}
\begin{eqnarray}\label{R42}
    C(\rho _{A \bar{B}}) = \frac{1}{\sqrt{e^{8 \pi (M - D)\omega} +1} } ,
\end{eqnarray}
and
\begin{eqnarray}\label{R43}
    C(\rho _{B \bar{B}}) = \frac{1}{\sqrt{e^{-8 \pi (M - D)\omega} + e^{8 \pi (M - D)\omega} + 2} } ,
\end{eqnarray}
respectively. It is found that, with the increase of the dilaton $D$, the physically accessible entanglement decreases monotonically, and at the same time, the physically inaccessible entanglement only increases monotonically. However, the physically inaccessible steering increases monotonically or non-monotonically with the growth of the dilaton $D$.
The  monogamous  relations between the fermionic steering and entanglement can be obtained as
\begin{eqnarray}\label{R44}
    S^{A \rightarrow B} - S^{A \rightarrow \bar{B} } = C^2(\rho _{AB}) - C^2(\rho _{A \bar{B}}),
\end{eqnarray}
\begin{eqnarray}\label{R45}
    S^{A \rightarrow B} + S^{A \rightarrow \bar{B} } = C^2(\rho _{AB}) + C^2(\rho _{A \bar{B}}) - \frac{2}{\sqrt{3} } C^2(\rho _{B \bar{B}}) ,
\end{eqnarray}
\begin{eqnarray}\label{R46}
   \frac{3 -\sqrt{3} }{2}  [S^{B \rightarrow A} - S^{\bar{B} \rightarrow A } ]= C^2(\rho _{AB}) - C^2(\rho _{A \bar{B}}) , (D > D_0),
\end{eqnarray}
\begin{eqnarray}\label{R46}
    \frac{3 + \sqrt{3} }{2}  [S^{B \rightarrow A} + S^{\bar{B} \rightarrow A } ]= C^2(\rho _{AB}) + C^2(\rho _{A \bar{B}}) , (D > D_0).
 \end{eqnarray}
These monogamous  relations show the relations between the fermionic entanglement and steering for physical accessibility and inaccessibility in dilaton spacetime.
They reflect that the  dilaton of the black hole can cause the transformations between these different types of quantum correlations.

\section{CONCLUSIONS \label{GSCDGE 5}}
In this paper, we have studied the distribution of the fermionic steering and the relation between fermionic Bell nonlocality, steering, and entanglement in the context of the dilaton black hole. Our model involves three observers: Alice, Bob, and Anti-Bob.
Here, Alice stays stationary at an asymptotically flat region, Bob hovers near the event horizon of the dilaton black hole, and Anti-Bob is restricted by the event horizon of the black hole.
We get that the fermionic steering between Alice and Bob decreases to a fixed value with the dilaton, while the bosonic steering suffers sudden death under the influence of the dilaton \cite{J49}.
In addition, the fermionic steerability from Alice to Bob is always larger than the fermionic steerability from Bob to Alice, whereas the bosonic steerability is the opposite of the fermionic steerability.
These  different properties between the fermionic and bosonic steering originate from the difference in statistics.
We find that the physically accessible steering in the extremely dilaton black hole cannot be affected by frequency. Interestingly, the physically accessible steering in the extremely dilaton black hole has no Bell nonlocality, meaning that quantum steering cannot be considered nonlocality.
We also find that quantum steering between Alice and Bob is a two-way steering in dilaton spacetime.

We also investigate the properties of the physically inaccessible steering in dilaton spacetime. It is shown that the dilaton gravity can redistribute the fermionic steering, but cannot redistribute Bell nonlocality, meaning that the physically inaccessible steering cannot be considered to be nonlocal. We find that, as the dilaton increases, the steering between Alice and Anti-Bob increases monotonously, while the steering from Bob to Anti-Bob increases non-monotonically.
When the steering from Anti-Bob to Alice experiences a sudden birth with the dilaton,
we obtain the maximum steering asymmetry that indicates the transformation between one-way steering and two-way steering in dilaton spacetime. When the steering from Bob to Anti-Bob experiences a sudden death with the dilaton, it shows the transformation between one-way  steering and no-way steering. Finally, we obtain some  monogamous relations between the fermionic steering and entanglement in dilaton spacetime. Therefore, we can indirectly understand the redistribution of quantum steering by understanding the redistribution of quantum entanglement in curved spacetime.

\begin{acknowledgments}
S.M. Wu was supported by the National Natural
Science Foundation of China (12205133) and LJKQZ20222315. J. Lu	was supported by National Natural Science Foundation of China (12175095) and  LiaoNing Revitalization Talents Program (XLYC2007047).
\end{acknowledgments}

$\textbf{Data Availability Statement}$

This manuscript has no associated data.

$\textbf{Conflict of interest}$

The authors declare no conflicts of interest.


\end{document}